\documentclass[aps,manuscript,showpacs,showkeys]{revtex4}
\usepackage{graphicx}
\usepackage{epsfig}  
\usepackage{epsf}    
\usepackage{dcolumn}
\usepackage{bm}
\usepackage{dcolumn}
\def\be{\begin{equation}}
\def\ee{\end{equation}}
\def\bea{\begin{eqnarray}}
\def\eea{\end{eqnarray}}
\def\gsim{\ \rlap{\raise 2pt\hbox{$>$}}{\lower 2pt \hbox{$\sim$}}\ }
\def\lsim{\ \rlap{\raise 2pt\hbox{$<$}}{\lower 2pt \hbox{$\sim$}}\ }
\def\dslash{\kern-4pt \not{\hbox{\kern-2pt $\partial$}}}
\def\pslash{\not{\hbox{\kern-2pt p}}}

\def\pmue{{{\rm P_{\mu e}} }}

\def\dcp{\delta_{CP}}

\begin{document}


\title{Magical properties of a 2540 km baseline Superbeam Experiment
 }


\author{Sushant K. Raut}
\affiliation{
Department of Physics, Indian Institute of Technology Bombay,
Mumbai 400076, India}

\author{Ravi Shanker Singh\footnote{Address after August 17, 2009: 
Department of Physics, Brown University, Providence, R.I., USA}}
\affiliation{
Department of Physics, Indian Institute of Technology Bombay,
Mumbai 400076, India}
 
\author{S. Uma Sankar\footnote{Corresponding Author}}
\affiliation{
Department of Physics, Indian Institute of Technology Bombay,
Mumbai 400076, India}
\date{\today}
\begin{abstract}
Lack of any information on the CP violating phase $\dcp$ 
weakens our ability to determine neutrino mass hierarchy. 
Magic baseline of 7500 km was proposed to
overcome this problem. However, to obtain large enough
fluxes, at this very long baseline, one needs new techniques
of generating high intensity neutrino beams. In this letter,
we highlight the {\it magical} properties of a 2540 km baseline.
At such a baseline, using a narrow band neutrino superbeam whose 
no oscillation event rate peaks around the energy 3.5 GeV, 
we can determine neutrino mass
hierarchy {\bf independently of} the CP phase. For $\sin^2 2 
\theta_{13} \geq 0.05$, a very modest exposure of 10 Kiloton-years
is sufficient to determine the hierarchy.
For $0.02 \leq \sin^2 2 \theta_{13} \leq 0.05$, an
exposure of about 100 Kiloton-years is needed.

\end{abstract}
\maketitle

\section{Introduction}

Neutrino experiments in the last decade have determined 
a number of neutrino parameters to a good accuracy. Among 
the currently unknown quantities are (i) the Chooz mixing
angle $\theta_{13}$, (ii) the sign of atmospheric mass-squared
difference and (iii) the CP violating phase $\dcp$. 
Experiments are being designed/constructed to measure these
quantities. 

At present, there are three efforts to measure a non-zero
value for $\theta_{13}$ using reactor neutrinos as the source.
In each of these experiments, the survival probability
$P(\bar{\nu}_e \to \bar{\nu}_e)$ will be measured using a pair
of identical detectors, one close to the reactor and the other  
about a kilometer away. The deficit
in the far detector is a measure of $\sin^2 2 \theta_{13}$.
Since these are all disappearance experiments, they should have
very low systematic uncertainties, to measure the small value
of $\sin^2 2 \theta_{13}$. Double Chooz will start taking 
data soon and it will see a positive signal if 
$\sin^2 2 \theta_{13} \geq 0.04$ \cite{dchooz}.
Daya Bay \cite{dayabay} and RENO \cite{reno} are expected to 
improve on this measurement.
Daya Bay's final sensitivity extends up to $\sin^2 2 \theta_{13} 
\geq 0.01$ \cite{lindner09}. 

It is possible to determine $\sin^2 2 \theta_{13}$ by measuring
$P(\nu_\mu \to \nu_e)$ ($\pmue$) in an accelerator experiment. 
T2K \cite{t2k} 
and NO$\nu$A \cite{nova} experiments aim to do this. Even if these 
experiments see a positive signal, determination of $\sin^2 2 \theta_{13}$
from their data will be subject to very large uncertainties 
because the probability $\pmue$ depends
on all three unknowns mentioned above \cite{lindner03}. 
The reactor experiments, on the other hand, will give a clean 
measurement of $\sin^2 2 \theta_{13}$ because 
$P(\bar{\nu}_e \to \bar{\nu}_e)$, at the relevant energies, 
depends only on this unknown neutrino parameter.

We label the three neutrino mass eigenstates by their respective 
eigenvalues $m_1$, $m_2$ and $m_3$. From the three masses, we can
define two independent mass-squared differences $\Delta_{21}
= m_2^2 - m_1^2$ and $\Delta_{31} = m_3^2 - m_1^2$. 
$\Delta_{21}$ is the mass-squared difference which drives
the solar neutrino oscillations. It is known to be positive
and its magnitude is much smaller than that of $\Delta_{31}$.
Atmospheric neutrino oscillations are essentially driven by 
$\Delta_{31}$ 
whose magnitude is known but not the sign.
If the neutrino masses follow the hierarchy $m_3 \gg m_2 > m_1$,
called normal hierarchy (NH), $\Delta_{31}$ is positive.  
It is negative if the neutrino masses have the pattern 
$m_2 > m_1 \gg m_3$, called inverted hierarchy (IH).
Present data allows both the possibilities.
Determining the sign of $\Delta_{31}$ establishes 
the pattern (or hierarchy) of neutrino masses. In this
letter, we propose a new scheme to realise this goal.

Neutrino propagation through dense matter, leads to an effective 
mass-squared term (usually called the matter term) in their Hamiltonian 
\cite{wolfenstein}.
The interference of this term with the original
mass-squared differences leads to the modification of the 
neutrino masses and mixing angles and hence their oscillation 
probabilities. The change induced by the interference, of course,
depends on the sign of the original mass-squared difference and
hence is different for NH and IH. By measuring this change 
in the oscillation probability $\pmue$, induced by the matter 
term, we can determine the mass hierarchy.

$\pmue$ depends on $\theta_{13}$, mass hierarchy and 
$\dcp$, all presently unknown. It is possible to obtain
the same $\pmue$ for various different combinations of 
these parameters \cite{degeneracy1,degeneracy2,magic}. It is desirable to 
design experiments which can measure each parameter 
individually.
Reactor neutrino experiments \cite{dchooz,dayabay,reno} can
make an unambiguous measurement of $\theta_{13}$. Daya Bay, in
particular, can make this measurement if $\sin^2 2 \theta_{13} > 0.01$
\cite{dayabay,lindner09}.
Such a measurement leads to disentanglement of one parameter. 
The CP violating phase $\delta_{CP}$, in principle, can be determined
by measuring the difference in oscillation probabilities of 
neutrinos and anti-neutrinos. However, this method cannot be used
to disentangle $\dcp$ and the matter term because   
the matter term changes sign under CP 
and induces a `CP violation-like' change in the probabilities. 
Hence a measurement of $\pmue$ will give two 
degenerate solutions, each with a different hierarchy and
a different CP phase. Various strategies have been proposed
to overcome this degeneracy problem \cite{degeneracy1,degeneracy2}. 

A radical proposal was made sometime ago to disentangle 
$\dcp$ from the matter term in $\pmue$.
The expression for $\pmue$ contains three terms and only  
one of them is dependent on $\dcp$.
If this term can be made to vanish, 
by an appropriate choice of neutrino energy and baseline, then 
it is possible to determine neutrino hierarchy without any 
information on $\dcp$. Calculation of the baseline, 
called {\it magic baseline}, gives an answer $L \simeq 7500$ km 
\cite{magic} and it is independent of energy \cite{smirnov}. 
Having sufficient neutrino
fluxes at such a large distance from the source will be very
difficult with the current accelerator technology. 

In this letter, we make an alternative proposal of a much 
shorter {\bf magical} baseline. In the original magic baseline
proposal, the condition that the $\dcp$ dependent term
vanish, holds both for NH and IH. This condition is quite
restrictive and leads to such a large baseline. We propose an
alternative condition which demands that the $\delta_{CP}$ terms
should vanish only for IH. This leads to a relation between 
the neutrino energy $E$ and the baseline $L$. In addition, 
we also demand
that $\pmue$ should be large for NH. This leads to
a different condition on $E$ and $L$. Solving these two 
equations gives us the solutions $L=2540$ km and $E=3.3$ GeV.
At this energy, for this baseline, $\pmue$
is very small for IH and near the maximum for NH, for any value of 
$\delta_{CP}$. Thus a neutrino beam, whose unoscillated 
event rate is maximum at this 
energy, can make a clean measurement of neutrino mass hierarchy,
independently of $\delta_{CP}$.  

\section{Calculation}

A very good approximate expression for $\pmue$, 
for three flavour oscillations including matter effects,
is usually given as an expansion in the small parameter
$\alpha = \Delta_{21}/\Delta_{31}$.
It can be written as \cite{numu2nue}
\begin{eqnarray}
\pmue & = & 
C_0 \frac{\sin^2 ((1-\hat{A}) \Delta)}{(1-\hat{A})^2} \nonumber \\
& + & \alpha \ C_1  \frac{\sin((1-\hat{A}) \Delta)}{(1-\hat{A})}
\ \frac{\sin(\hat{A}\Delta)}{\hat{A}} \nonumber \\
& + & \alpha^2 \ C_2 \frac{\sin^2(\hat{A}\Delta)}{\hat{A}^2},
\label{pmue}
\end{eqnarray} 
where $\Delta = (1.27 \Delta_{31} L/E)$ and $\hat{A} = A/\Delta_{31}$. 
The matter term \cite{wolfenstein} $A~({\rm in~eV^2})
= 0.76 \times 10^{-4} \rho~({\rm gm/cc})~E~({\rm GeV})$. 
$\rho$ is the density of matter through which the neutrino propagates. 
Here $\Delta_{31}$ is 
given in units of eV$^2$, $L$ is in km and $E$ is in GeV. 
The coefficients, $C_i$ are given by 
\begin{eqnarray}
C_0 & = & \sin^2 \theta_{23} \sin^2 2 \theta_{13} \\
C_1 & = & \cos \theta_{13} \sin 2 \theta_{12} \sin 2 \theta_{13} 
\sin 2 \theta_{23} \cos (\Delta+\delta_{CP}) \\
C_2 & = & \sin^2 2 \theta_{12} \cos^2 \theta_{23} 
\end{eqnarray} 
We note that only $C_1$ among them depends on the phase $\delta_{CP}$.  

$\pmue$ depends on all the three unknowns. Data on $\pmue$
from a single experiment leads to a degenerate set
of solutions \cite{barger02}. Data on $\pmue$ from experiments
with different baselines can resolve some of these degeneracies
\cite{twobaseline}. Here we assume that $\theta_{13}$ 
will be measured in reactor neutrino experiments 
\cite{dchooz,dayabay,reno}, which 
resolves the degeneracies involving this parameter. Our proposal 
in this letter makes the hierarchy-$\dcp$ degeneracy irrelevant. 

In Eq.~(\ref{pmue}) the dependence on the matter term $\hat{A}$ is 
explicitly displayed. $\Delta_{31}$ is positive for NH and is 
negative for IH. $A$, on the other hand, is positive for neutrinos 
and is negative for anti-neutrinos. Thus, if we have only a neutrino 
beam, then $\hat{A}$ is positive for NH and is negative for IH. 
For anti-neutrino beam, the situation is reversed. In Eq.~(\ref{pmue}), 
the first term is the most sensitive to hierarchy but the second
term provides a significant correction. As one varies $\delta_{CP}$
in its range, the change in the second term can cancel the
change in the first term caused by the change in hierarchy.
In other words, there exist two degenerate sets of solutions,
(NH and $\dcp = \delta_1$) and (IH and $\dcp = \delta_2$),
both of which give the same value of $\pmue$ in a given experiment
\cite{degeneracy1,barger02}. 
This leads to hierarchy-$\dcp$ degeneracy and restricts our ability 
to determine the hierarchy.
To overcome this, it was proposed to choose 
a baseline and energy for which $\sin (\hat{A} \Delta) = 0$, so 
that the second and third terms vanish. The above constraint, for
the first non-trivial zero, gives the magic baseline condition
$L \approx 7500$ km, independent of the energy \cite{magic}. The energy can be 
chosen by the condition that the oscillation probability be a 
maximum at this baseline. Therefore, one can indeed determine mass 
hierarchy at the magic baseline in a clean manner, independently of 
$\delta_{CP}$. However, one is now faced with the problem of 
obtaining large enough flux at this large a distance. 

We can make $\pmue$ independent of $\dcp$ by choosing either 
$\sin (\hat{A} \Delta) = 0$ or $\sin ((1-\hat{A}) \Delta) = 0$
\cite{smirnov}.
The magic baseline uses the first condition which holds 
true for both NH and IH for the same $L$ and is independent of $E$. 
The dependence of the second condition on $L$ and $E$ is 
different for NH and IH. We exploit this difference by choosing
$L$ and $E$ such that $\sin ((1-\hat{A}) \Delta) = 0$ holds for IH.
This makes $\pmue$ for IH not only independent of $\dcp$ but also
very small because only the $\alpha^2$ term in Eq.~(\ref{pmue}) 
survives. We impose the simultaneous demand that, for the same
$L$ and $E$, $\pmue$ for NH should be close to maximum. This leads
to a substantial difference in $\pmue$ for NH and IH and enables
us to determine the neutrino mass hierarchy for all values of $\dcp$,
even for relatively small values of $\theta_{13}$. 
The condition on $\pmue$ for IH translates into 
$1.27 (|\Delta_{31}| + A) L/E = \pi$ whereas that for NH 
becomes $1.27 (|\Delta_{31}| - A) L/E = \pi/2$.
Solving the above two equations, we get $L = 2540$ km and 
$E=3.3$ GeV, for $|\Delta_{31}| = 2.5 \times 10^{-3}$ eV$^2$
\cite{minos}. 

For later convenience,
we define the product of the neutrino flux with 
cross section to be "no oscillation event rate (NOER)".
The event rate is then a product of NOER and the oscillation
probability. For $\nu_\mu \rightarrow \nu_e$, this probability 
largely depends on 
$\sin^2[1.27(|\Delta_{31}|\pm A)L/E]$. The three quantities 
$|\Delta_{31}|$, $A$ and $E$ all contain uncertainties of order a 
few percent. Thus the phase $[1.27(|\Delta_{31}|\pm A)L/E]$ is uncertain by a 
few percent. However, for the chosen values of $L$ and $E$,  
the square of the sine of this phase
is close to its maximum or minimum. Hence an
uncertainty of a few percent in this phase will lead to a much 
smaller uncertainty in the square of the sine function 
and hence on the event rate.

The large difference between NH and IH is illustrated in Fig.~1, 
where we have plotted $\pmue$ as
a function of $E$ for $\sin^2 2 \theta_{13} = 0.02$. The figure
shows $\pmue$ as a band with $\dcp$ spanning the entire 
$0^\circ-360^\circ$ range, for both NH and IH. As claimed above, 
in the neighbourhood of $E=3.3$ GeV, $\pmue$ is about $0.002$ for all 
$\dcp$ for IH whereas it is $6$ to $18$ times larger 
(depending on $\dcp$) in the case of NH.
This large difference can be measured if there is a substantial 
NOER at $E=3.3$ GeV.  
Note that
the value of $\sin^2 2 \theta_{13}$ chosen here is quite small and is
barely above the detectable limit of the experiments under construction.
But even for such small values of $\theta_{13}$, the configuration
suggested here can make a distinction between NH and IH, 
{\it independently of $\dcp$ and using neutrino beam only}. 
It is worth remarking that $2540$ km is the distance 
between the Brookhaven Laboratory and the Homestake Mine 
\cite{brookhome}.
The design for a neutrino superbeam, with NOER peaking near $3.3$ GeV,
already exists. For example: NuMI beam with medium energy option 
at locations $7$ mr off-axis, has its peak NOER at $3.5$ GeV \cite{nova}.
\begin{figure}
\begin{center}
\epsfig{file=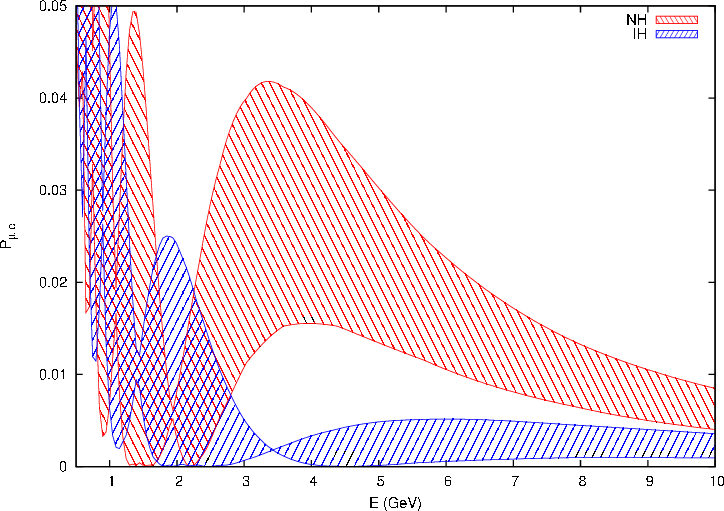,width=6in}
\caption{\footnotesize
$\pmue$ as function of $E$ for $L=2540$ km and $\sin^2 2 \theta_{13} = 0.02$. 
It is plotted for both NH and IH, in each case as a band. Within each band,
$\dcp$ varies in the range $0^\circ-360^\circ$.}
\label{fig1}
\end{center}
 \end{figure}


In this letter, we consider the following configuration. We
assume that the $\nu_\mu$ source is located at Brookhaven which produces
a NuMI-like beam with $0.8$ megawatt beam power. This corresponds 
to $7.3 \times 10^{20}$ protons on target (POT), with energy 120 GeV, 
per year \cite{nova}. We assume a
$300$ Kt water Cerenkov detector at Homestake $2540$ km away.
We also assume that the orientation of the beamline is such 
that the detector location is $7$ mr off-axis. 
In our calculations, we take the NuMI NOERs for this off-axis location 
and scale them appropriately to obtain
the NOERs at this distance
\cite{messier}. Our signal is electron appearance in the far detector,
due to $\nu_\mu \to \nu_e$ oscillations.  
Single $\pi^0$ events produced by neutral current interactions form a
potentially huge background to this signal. The visible energy of the 
neutral current events is usually much smaller than the true energy 
of the neutrino and hence they can be suppressed by a large factor 
\cite{brookhome}. The remainder
of these neutral current events, together with the electron events 
produced by beam $\nu_e$ form the actual background. This background
was estimated in a previous study to be about $1\%$ 
of the unoscillated events \cite{brookhome}. 
We include this background in our study. 
We calculate the number of events for the energy range $1-10$ GeV, 
in bins of width $0.4$ GeV and smear the obtained event distribution 
with a Gaussian probability in energy with $\sigma_E = 0.15E$.
We define the statistical $\chi^2$ between the event distribution for NH
and that for IH, by 
\begin{equation}
\chi^2_{stat} = \sum_{i=bins} \frac{(N^{TR}_i - N^{TE}_i)^2}{N^{TR}_i}.
\end{equation}
$N^{TR}_i$ is the number of events for the {\bf true}
hierarchy plus the number of background events. 
$N^{TE}_i$ is the number of events for the {\bf test}
hierarchy, which is the opposite of the true hierarchy,
plus the number of background events. 
The true hierarchy  
can be either NH or IH and we consider both possibilities. 
In calculating the event distributions, the following
values of neutrino parameters are used: $\Delta_{21} = 8 \times 
10^{-5}$ eV$^2$, $|\Delta_{31}| = 2.5 \times 10^{-3}$ eV$^2$,
$\sin^2 \theta_{12} = 0.31$ and $\sin^2 \theta_{23} = 0.5$
\cite{globalfits}.
We do the calculation for various different input values
of $\sin^2 2 \theta_{13}$ starting from $0.1$ and going down
to $0.02$ in steps of $0.01$. $\dcp$ is varied from $0$
to $360^\circ$ in steps of $45^\circ$. 

We also assume a $2\%$ systematic uncertainty in the neutrino flux and 
a similar uncertainty in detector systematics. The systematic 
uncertainty in the cross section is taken to be $10\%$ \cite{nova}.
These are taken into account through the method of pulls
as described in \cite{fogli1,fogli2,gg}.
In this method, the fluxes, cross sections, etc are 
taken to be the central values in the computation
of $N^{TR}_i$ but are allowed 
to deviate from their central values in the computation of
$N^{TE}_i$. We assume that the $k^{th}$ input deviates from 
its central value by $\sigma_k \;\xi_k$,
where $\sigma_k$ is the uncertainty in this input. Then the value of
$N^{TE}_i$ with the changed inputs is given by
\begin{equation}
N^{TE}_i =  N^{TE}_i(std) + \sum^{npull}_{k=1}\;
c_i^k \;\xi_k 
\label{cij}
\end{equation}
where $N^{TE}_i(std)$ is the expected number of events in  
bin $i$ for the {\bf test} hierarchy, calculated using the 
central values of the fluxes, cross sections etc and npull 
is the number of inputs which have systematic uncertainties.
The $\xi_k$'s are called the {\it pull} variables and they
determine the number of $\sigma$'s by which the
$k^{th}$ input deviates from its central value. In
eq.~(\ref{cij}), $c_i^k$ is the change in $N^{TE}_i$
when the $k^{th}$ input is changed by $\sigma_k$ 
({\it i.e.}~by 1 standard deviation). The
uncertainties in the inputs are not very large. Therefore, in
eq.~(\ref{cij}) we consider only those changes in $N^{TE}_i$
which are linear in $\xi_k$. Thus we have a modified $\chi^2$ defined by
\begin{equation}
\chi^2(\xi_k) = \sum_i\;
\frac{\left[~N_i^{TE}(std) \;+\; \sum^{npull}_{k=1}\; c_i^k\;
\xi_k - N_i^{TR}~\right]^2}{N_i^{TR}} + \sum^{npull}_{k=1}\;
\xi_k^2  
\label{chisqxik}
\end{equation}
where the additional term $\xi_k^2$ is the penalty imposed
for moving $k^{th}$ input away from its central value by
$\sigma_k \;\xi_k$. The $\chi^2$ with pulls, which includes
the effects of all theoretical and systematic uncertainties, is
obtained by minimizing $\chi^2(\xi_k)$, given in
eq.~(\ref{chisqxik}), with respect to all the pulls $\xi_k$:
\begin{equation}
\chi^2_{pull} = Min_{\xi_k}~ \left[~ \chi^2(\xi_k)~\right].
\label{chisqpull}
\end{equation}

In addition to taking the systematic uncertainties into account, 
we have marginalized over $|\Delta_{31}|$, $\sin^2 2 \theta_{13}$, 
$\sin^2 2 \theta_{23}$ and $\dcp$ but held $\Delta_{21}$ and 
$\theta_{12}$ fixed. In doing the marginalization, we assume that 
experimental uncertainties in $|\Delta_{31}|$ and $\sin^2 2 
\theta_{23}$ are what they are expected to be from T2K (about $2\%$)
\cite{t2k}. 
It turns out that the marginalization over $|\Delta_{31}|$ has
on $\chi^2$ no effect but marginalization over $\theta_{23}$ has a 
very significant effect, if IH happens to be the true hierarchy. 
We elucidate this point after discussing our results.

The dominant term in $\pmue$ is proportional to 
$\sin^2 2 \theta_{13}$, which makes the marginalization  
over $\theta_{13}$ the most crucial one in hierarchy determination.
In the neighbourhood 
of oscillation maximum, the matter effects increase this term for NH 
and decrease it for IH, relative to its vacuum value. In shorter
baseline experiments such as NO$\nu$A, one expects to measure this
increase/decrease and determine the hierarchy. However, it is 
possible to choose a $\theta_{13}^{'}$, within the allowed range 
of $\theta_{13}$, such that $\pmue({\rm NH},\theta_{13}) \simeq
\pmue({\rm IH},\theta_{13}^{'})$ \cite{barger02}. 
In such a situation, marginalization
over $\theta_{13}$ leads to a very small $\chi^2$. In our proposal,
the condition $\sin ((1-\hat{A}) \Delta) = 0$, makes $\pmue$(IH) 
very small, in the neighbourhood of $E=3.3$ GeV, {\bf independently 
of both $\dcp$ and $\theta_{13}$}. In this energy range, $\pmue$(NH) 
is close to oscillation maximum. It is also
proportional to $\sin^2 2 \theta_{13}$ and 
is quite large even for the very small value of $\sin^2 2 \theta_{13}$,
which is illustrated in Fig.~1. Therefore the oscillation pattern is 
very distinctive for each hierarchy and they can easily be distinguished
for $\sin^2 2 \theta_{13} \geq 0.02$. We also note that
the marginalization over $\dcp$ ensures that this distinction between
NH and IH exists even for the most unfavourable value of $\dcp$. 
Hence hierarchy determination is possible for the {\bf whole range of
$\dcp$.}

Because of its importance, marginalization over 
$\sin^2 2 \theta_{13}$ was done in different stages.
At present, we only have an upper limit, $\sin^2 2 \theta_{13} \leq 0.1$.
In this study, we will restrict ourselves to the values
$\sin^2 2 \theta_{13}$ accessible to the experiments
currently running or under construction. 
Double Chooz \cite{dchooz} can measure this parameter if it is $\geq 0.04$
and Daya Bay \cite{dayabay} can measure it for values $\geq 0.01$. 
Therefore, for 
input values $\sin^2 2 \theta_{13}: 0.05 - 0.1$ (that is if Double Chooz
finds a positive result), we do marginalization only over this range.
For input values $\sin^2 2 \theta_{13}:
0.02 - 0.05$ (expecting a positive result from Daya Bay but not from Double
Chooz) we do marginalization only over this restricted range. 

It was mentioned above that the parameters $\Delta_{21}$ and
$\theta_{12}$ are kept fixed. These parameters occur only in
the second and third terms of $\pmue$. Given the smallness 
of $\alpha$, the third term is very small and varying 
$\Delta_{21}$ and $\theta_{12}$ within their ranges, changes this
term by a very small amount. The second term undergoes much larger changes 
when $\sin 2 \theta_{13}$ and $\dcp$ are varied over their ranges.
The change due to $\Delta_{21}$ and $\theta_{12}$ is much smaller
and can be neglected. 

\section{Results}

In our calculations, we assumed that only the neutrino beam is
used. Neutrino beams have an advantage over the anti-neutrino
beams because of the larger cross section and hence larger
statistics. In table~1, we list the exposure, in Kiloton-years
(Kt-yr), 
needed to obtain a $3 \sigma$ distinction between NH and IH if 
$\sin^2 2 \theta_{13} \geq 0.05$.  
In this case, one requires only minimal
exposure (less than 10 Kt-yr) to distinguish the two hierarchies. 
This is independent of whether the true hierarchy is NH or IH.
For $0.05 \geq \sin^2 2 \theta_{13} \geq 0.02$, 
the results are shown
in table~2. Here the needed exposure, for $3 \sigma$ distinction, 
sharply goes up with $\theta_{13}$ from $7$ Kt-yr to $50$ Kt-yr, 
if NH is the true hierarchy. If IH is the true hierarchy, the 
$3 \sigma$ exposure becomes about $92$ Kt-yr.  

\begin{table}[tbh]
\begin{tabular}{|c|c|c|}
\hline

$\sin^2 2\theta_{13}$ (true) & Exposure(NH) & Exposure(IH) \\ \hline
0.1 & 2.93  & 6.39 \\
 \hline
0.09 & 3.34 & 6.04 \\
 \hline
0.08 & 3.94 & 5.69 \\
 \hline
0.07 & 4.77 & 5.38 \\
 \hline
0.06 & 6.04 & 4.95 \\
 \hline
0.05 & 8.19 & 4.55 \\
 \hline
\end{tabular}
\caption{Exposure in Kiloton-years required for $3\sigma$ hierarchy 
discrimination in the case where both Double Chooz
and Daya Bay see a positive signal. Second (third) column shows the 
results if NH (IH) is the true hierarchy.}
\label{DC_DB}
\end{table}

\begin{table}[tbh]
\begin{tabular}{|c|c|c|}
\hline

$\sin^{2}2\theta_{13}$ (true) & Exposure(NH) & Exposure(IH) \\ \hline
0.05 & 7.32 & 91.68\\
 \hline
0.04 & 10.69 & 86.80 \\
 \hline
0.03 & 18.97 & 81.16 \\
 \hline
0.02 & 49.83 & 71.50 \\
 \hline
\end{tabular}
\caption{Exposure in Kiloton-years required for $3\sigma$ 
hierarchy discrimination
in the case where Daya Bay shows a positive signal, 
but Double Chooz does not.}
\label{DB}
\end{table}

Note that the exposures, for NH being the true hierarchy, 
increase with decreasing values of $\sin^2 2 \theta_{13}$ 
whereas, if IH is the true hierarchy, the exposures are 
more or less independent of the true value of 
$\theta_{13}$. This feature occurs due to the marginalization
over $\theta_{13}$. The number of events in the region of peak 
NOER of $3.5$ GeV are strongly dependent on $\theta_{13}$ for NH, whereas 
they are independent of $\theta_{13}$ for IH, as can be seen 
from eq.~(\ref{pmue}).  
If NH is the true hierarchy, $N^{NH}_i$ is
computed using the input value of $\theta_{13}$ but in computing
$N^{IH}_i$ we vary $\theta_{13}$ in its marginalizing range. 
Since the difference between these two numbers, decreases with 
decreasing $\theta_{13}$, we need larger exposure to obtain the
same $\chi^2_{min}$. If IH is the true hierarchy, then $N^{IH}_i$ is
computed using the input value of $\theta_{13}$ and $N^{NH}_i$
is computed with varying values of $\theta_{13}$. 
Around 3.5 GeV. $N^{IH}_i$ is small and independent of $\theta_{13}$,
whereas, $N^{NH}_i$ takes its smallest value for the smallest of 
the $\theta_{13}$ values in the marginalizing range. Therefore 
$\chi^2_{min}$ and the exposure for $3 \sigma$ distinction, 
have a much weaker dependence on the input value of $\theta_{13}$, 
if IH is the true hierarchy.
The small decrease in the exposure with decreasing values of
$\theta_{13}$ occurs due to the contribution of events in the 
energy range beyond $4$ GeV. For this range, the difference 
between $N_i^{IH}$ and $N^{NH}_i$ increases as the input value 
of $\theta_{13}$ is decreased. Thus $\chi^2_{min}$ increases 
and exposure decreases. 

For $0.02 \leq \sin^2 2 \theta_{13} \leq 0.05 $, the exposure
is rather large if IH is the true hierarchy. This occurs due 
to the marginalization over $\theta_{23}$. Without marginalization
over this parameter, the required exposure is about $50$ Kt-yr 
which is similar to the largest exposure needed if NH is the 
true hierarchy. With marginalization, the IH spectrum is computed
with $\theta_{23}$ equal to the input value of $\pi/4$ whereas
the NH spectrum is computed with different values of $\theta_{23}$
in the allowed range \cite{t2k}. 
Given that the number of events in
the case of IH is very small, the difference between the IH spectrum 
and NH spectrum is smaller, when the NH spectrum is computed with
a smaller value of $\theta_{23}$. Thus $\chi^2_{min}$ will be smaller
and we need larger exposure to obtain a $3 \sigma$ discrimination.

The baseline 2540 km and the energy of peak NOER 3.3 GeV were obtained 
based on the two conditions that $\pmue$ for IH should be independent
of $\dcp$ (which makes it very small)  
and for NH it should be close to maximum. This leads to a large 
difference between the expected number of signal events for NH and IH.
However, the fluxes fall off as $1/L^2$ with distance. Therefore 
it is imperative to check that the baseline 2540 km is the optimum
distance to determine mass hierarchy with a NuMI-like neutrino
source in the medium energy option. The event rate is a product of flux, 
oscillation probability and the cross section. If $L$ is larger,
the maximum value of the oscillation probability  
occurs for higher values of $E$.  This leads
to a larger matter effect (and hence better separation between 
hierarchies) and to larger cross sections. 
Therefore, for each $L$,
we choose an off-axis angle such that the energy of peak NOER 
coincides with the energy of oscillation maximum for NH. These
energies and off-axis angles are listed in table~3, for some 
baselines varying between 1000-3000 km. Using the programs
available at \cite{messier}, we calculated the fluxes
for these baselines at the corresponding off-axis angles. 
Using these fluxes, we computed the $\chi^2_{min}$
for hierarchy discrimination with the following inputs: 
(i) an exposure of 150 Kt-yr, (ii) the true value of
$\sin^2 2 \theta_{13} = 0.02$ and (iii) the true hierarchy is NH.
The other parameter values and systematics are kept the
same as in the earlier calculations. 
The results are shown in table~3.
We note that the largest $\chi^2_{min}$ occurs 
for 2540 km baseline, showing that this indeed is the optimum distance,
for hierarchy discrimination, for the given neutrino source.
For shorter baselines, the off-axis angle is
larger, which makes the flux much smaller \cite{nova}.
Hence, the hierarchy discrimination ability of shorter baselines,
with this source, is much worse. 

\begin{table}[tbh]
\begin{tabular}{|c|c|c|c|}
\hline

Baseline & Peak Energy & Off-axis Angle 
& ~~$\chi^2_{min}$~~ \\ 
(in km) & (in GeV) & (in mr) &   \\ \hline
1000 & 1.5 & 20 & 0.06 \\
 \hline
1500 & 2.3 & 12.5 & 0.6 \\
 \hline
2000 & 3.0 & 10 & 5.9  \\
 \hline
2540 & 3.5 & 7 & 26.2 \\
 \hline
3000 & 4.2 & 6 & 19.7 \\
 \hline
\end{tabular}
\caption{Baseline length $L$ vs $\chi^2_{min}$ for hierarchy 
discrimination for $\sin^2 2 \theta_{13} = 0.02$ with a NuMI-like
source with medium energy option for 150 Kt-yr exposure. Also 
given are the energy where $P_{\mu e}$ peaks and the corresponding
off-axis angle such that NOER peaks at the same energy. 
}
\label{chisq}
\end{table}

We have highlighted some very interesting properties
of a 2540 km baseline experiment and showed, through a simple
numerical calculation, that such an experiment is well capable
of determining neutrino mass hierarchy. In our calculations, 
we have included a conservative estimate of background, which
we take to be $1\%$ of the unoscillated events. Despite such 
background, the setup we discussed is capable of hierarchy 
discrimination for even quite small values of $\theta_{13}$.
By imposing various kinematic cuts, the background can be 
suppressed at the cost of loss of some signal. This loss of signal 
can be compensated by having an increased exposure. However,
the ability of the setup to determine the mass hierarchy 
will not be compromised because any such kinematic cut will lead
to a larger signal to background ratio. 

\section{Conclusion}

In this letter we demonstrated the superior ability of a 
neutrino superbeam experiment with a baseline 2540 km, 
whose NOER peaks in the energy range 3-4 GeV, to determine
neutrino mass hierarchy. For $\sin^2 2 \theta_{13} \geq 0.05$,
a very modest exposure of $\leq 10$ Kt-yr is sufficient
to distinguish the two hierarchies at $3 \sigma$ level. For
$0.02 \leq \sin^2 2 \theta_{13} \leq 0.05$, one needs an exposure
$\leq 100$ Kt-yr. These exposures are obtained analyzing 
the expected data from this superbeam set up only. If the data
from this set up is analyzed in conjunction with the data from
a reactor $\theta_{13}$ measurement experiment, then the required
exposures are likely to be much less. The set up we assumed is not 
hard to realize because 2540 km is the distance from Brookhaven
to Homestake and the technology for an accelerator beam, with
peak NOER in 3-4 GeV range, exists \cite{nova}. 
Such a set up, we believe, will have an excellent capability to measure 
not only small values of $\theta_{13}$ but $\dcp$ as well. These issues 
are currently being studied.

\noindent
{\bf Acknowledgement} Ravi Shanker Singh thanks BRNS project
and Prof. Asmita Mukherjee for financial support. We thank 
Srubabati Goswami for discussions regarding this problem and
Raj Gandhi for a critical reading of the manuscript.

\end{document}